\documentstyle[12pt,a4]{article}
\begin{document}
\normalbaselineskip=24 true pt
\normalbaselines
\addtolength{\topmargin}{-2cm}
\newcommand{\bm}{\bibitem}
\newcommand{\ud}{\bf}
\textheight=23 cm
\thispagestyle{empty}
\renewcommand{\thefootnote}{\fnsymbol{footnote}}
\begin{center}
{\Large $NN \rightarrow NN \pi$ reaction near threshold in a
covariant one-boson-exchange model
\footnote{Work supported by KFA J\"ulich and GSI Darmstadt}}
\\[1.0 cm]
{\bf{R. Shyam$^{(1)}$ and U. Mosel$^{(2)}$}\\
$^{(1)}${\it Saha Institute of Nuclear Physics, 1/AF Bidhan Nagar, Calcutta -
700064, India}}\\ [0.2 cm]
$^{(2)}${\it Institute f\"ur Theoretische Physik, Universit\"at Giessen,
D-35392 Giessen, Germany}\\
\end{center}
\renewcommand{\thefootnote}{\arabic{footnote}}
\begin{abstract}
We calculate the cross sections for the $p(p,n\pi^{+})p$ and $p(p,p\pi^{0})p$
reactions for proton beam energies near threshold in
a covariant one-boson-exchange model, which incorporates the
exchange of $\pi$, $\rho$, $\sigma$ and $\omega$ mesons and treats both
nucleon and delta isobar as intermediate states. The final state interaction
effects have also been included. The $\omega$ meson exchange is found to
contribute significantly at these energies, which, along with
other meson exchanges, provides an excellent agreement with the data.
The cross sections at beam energies $\leq$ 300 MeV are almost free from the
contributions of the $\Delta$ isobar excitation.


\end{abstract}
\newpage
High precision data on the total as well as differential cross
sections for the $p(p,p\pi^{0})p$ and $p(p,n\pi^{+})p$ reactions at the
incident proton energies very close to the kinematical threshold have now
become available [1,2,3]. At low energies, these reactions necessarily
involve large momentum transfers, which make them very sensitive to
the nucleon-nucleons (NN) interactions at small distances.
These data provide a stringent test of the theories used to interpret
pion production in elementary NN collisions.

Earlier attempts
[4,5] to explain the $\pi^{0}$ data in the
$s$-wave pion production approach of Koltun and Reitan [6], have not been
successful as the calculations underpredicted the data by factors of 4-5.
More recently, Lee and Riska [7] and Horowitz et al. [8] have shown that
most of the theoretical underprediction found in earlier calculations,
are removed if the exchange of heavy scalar and vector mesons is
included in the calculations. However, both these sets of authors
have difficulty (of varying degree) in
explaining the energy dependence of the measured cross sections. An
alternative explanation of the problem has been presented by Hernandez and
Oset [9], who argue that if the off-shell properties of the pion-nucleon
($\pi N$) amplitude (which, in general, is much larger than its on-shell
value) are taken into account explicitly while evaluating the pion-
rescattering diagram, the calculated $p(p,p\pi^{0})p$
cross sections come very close to the measured values. However,
Hanhart et al. [10] show that with the
off-shell $\pi N$ amplitude calculated properly, the enhancement of the
cross sections is much smaller than that reported in Ref. [9].

In the case of the $p(p,n\pi^{+})p$ reaction, the
models of Schillaci, Silbar and Young (SSY) [11],
and Lee and Matsuyama (LM) [12] have been used to explain the data.
SSY employ the partial conservation of axial vector current (PCAC) amplitude
for the pion production with several approximations to reduce the
complexity of the calculations. In the LM approach, the pion production
via the $\Delta$ isobar excitation is treated rather rigorously
within a coupled channel formalism, while the nonresonant
production process (which is important at low energies)
is handled very approximately.
Both these models are found to be inadequate to account for the
data [3].

A proper theoretical understanding of the near threshold $\pi^{0}$
and $\pi^{+}$ production in $pp$ collisons within one consistent picture
is still lacking. The exchange of heavy scalar and vector mesons is
important for the description of the short range part of the
NN interaction [13], which should be included in the calculation
of the pion production in NN collisions at low energies, in
a fully covariant way.
Most of the models mentioned above perform calculations in the
non-relativistic framework where inaccuracies creep in also
due to the ambiguity that exists in the non-relativistic
reduction of the $\pi NN$ Lagrangian [14].

In a recent publication [15], we have presented, in detail,
a fully covariant effective one-boson-exchange model (CEOBEM)
to describe the pion production in NN collisions.
CEOBEM incorporates the exchange of $\pi$, $\rho$,
$\sigma$ and $\omega$ mesons and treats both nucleon and delta isobar
excitations as intermediate states. The model involves only the physical
parameters (like, coupling constants and cutoff masses), which are
determined by fitting to the NN scattering data over a range of beam
energies. Since we directly fit the NN scattering T-matrix, we have
also included in CEOBEM a nucleon-nucleon-axial-vector-isovector meson
vertex. This term provides an additional short range correlation which
cures the unphysical behaviour in the angular distribution of NN
scattering caused by the contact term in the one-pion exchange
amplitude [15,16] in the limit of very large mass of the axial vector
meson.

Within CEOBEM, it is possible to describe both the $p(p,p\pi^{0})p$
and $p(p,n\pi^{+})p$ reactions at beam energies near the
threshold in one consistent picture, which is the aim of this letter.
For applications at low energies, the model as presented in
Ref. [15], has to be extended to include the final state
interaction (FSI) effects in the outgoing channel.

In the spirit of the Watson-Migdal theory of FSI [17], the
transition amplitude for a $NN \rightarrow NN \pi$ type of reaction
can be written as
\begin{eqnarray}
A_{fi} & = & T_{fi}(NN \rightarrow NN \pi)T_{ff}, \nonumber
\end{eqnarray}
where $T_{fi}(NN \rightarrow NN \pi)$ is the primary production amplitude
which describes the transition from initial $NN$ state to the final $NN \pi$
state, while the amplitude $T_{ff}$ accounts for the rescattering among
the final particles.
The essential approximation involved in this equation is the assumption
that all the diagrams that lead to the pion production are contained
in $T_{fi}$, while $T_{ff}$ takes into account the elastic scattering
among the final particles once they are produced by the processes involved
in $T_{fi}$. Such a factorization of the transition amplitude,
which can be justified under certain conditions by
the general scattering theory [18], has been found to provide a good
description of the near threshold $NN \rightarrow NN \eta$ reaction data [19].

We calculate the amplitude $T_{fi}$ within the CEOBEM, summing coherently the
contributions of various pion production diagramms in precisely
the same way as discussed in Ref. [15], with all the parameters being the
same as those given therein. On the other hand,
the evaluation of the rescattering amplitude $T_{ff}$ requires the
solution of a set of integral equations involving two-body interactions
among the different final state particles. However,
we simplify the problem by identifying $T_{ff}$ with the coherent sum of the
two-body on-mass-shell elastic scattering amplitudes of the particles
involved in the final channel [19]. Furthermore, we assume the
pion-rescattering terms to be negligible. Indeed, several authors
[7,8,10,20] agree with the fact that
the on-shell part of this term is close to zero; the off-shell part of the
$\pi N$ rescattering is assumed to be contained in the $\pi NN$ form
factors [15]. Under these conditions, the $T_{ff}$ is given by
the inverse of the Jost function, $J_{\ell}(k)$, for a given
partial wave $\ell$ and the relative momentum $k$ of the two
outgoing nucleons [18,21].

The function $J_{\ell}(k)$ can be obtained by solving the
Schr\"odinger equation with a given NN interaction
with proper boundary conditions.  As the FSI effects are most
important at low relative energies, one can safely ignore the partial
waves other that $\ell = 0$ [22] and (in the absence of Coulomb force)
can make the following effective range expansion for the $s$-wave
NN scattering phase-shift
\begin{eqnarray}
kcot\delta_{0} & = & \frac{1}{a_{0}} + \frac{1}{2}r_{0}k^{2}, \nonumber
\end{eqnarray}
where $a_{0}$ and $r_{0}$ are the scattering length and effective range
parameters respectively. An analogous expression can be
written also in the presence of the Coulomb interaction  (see eg. Ref. [18]).
Now, the inverse of the Jost function can be written as
\begin{eqnarray}
(J_{0}(k))^{-1} & = & \frac{(k^2 + \alpha ^2)r_{0}/2}
                         {1/a_{0} + (r_{0}/2)k^2 - ik}, \nonumber
\end{eqnarray}
where $\alpha$ is given by
\begin{eqnarray}
\alpha & = & (1/r_{0})[1 +  (1 + 2r_{0}/a_{0})^{1/2}]. \nonumber
\end{eqnarray}
The quantity $(J_{0}(k))^{-1}$ as written above, has the required property
that it goes to unity for large $k$. This is, of course, to be expected
since the final state interaction effects should disappear for large relative
energies of the two outgoing nucleons.
We have checked the accuracy of the Jost functions calculated in this way
by comparing them with those obtained by solving the
Schr\"odinger equation with the relevant parts of the Paris potential in
some cases. There is a good general agreement between the two results.
In case of the neutron-proton final channel, we have included both
$^{1}S_{0}$ and $^{3}S_{1}$ FSI's, while for the proton-proton
case only the former is considered. The corresponding scattering length
and effective range parameters have been taken from Ref. [23].

In Fig. 1, we compare the results of our calculations (solid line)
with the data for the total cross section of the $p(p,n\pi^{+})p$ reaction
as a function of the beam energy.  We see that the agreement between
calculations and the data is truly remarkable. Not only the absolute
magnitude but also the energy dependence of the experimetal cross sections
are well reproduced. Since our model includes the pion production via
both the nucleon and $\Delta$ isobar intermediate states, it can be extended
to higher beam energies in a straightforward way. We obtain a good
agreement with the data also at higher energies (E$_{lab}$ $>$ 320 MeV).
This is an unique feature of our calculations, as most of the recent models
used to explain the near threshold pion production lack the treatment of
both the non-resonant as well as resonant pion production
consistently in a single framework. Therefore, the failure of these models
to describe the data both at low and high beam energies is understandable.

In this figure, we also show the contributions of the various
meson-exchange graphs to the total pion production cross sections. It is
clear that the contribution of $\omega$ meson exchange is quite significant;
very close to the threshold it is even larger than that of
the pion exchange. In comparison to this, the strength of
the scalar $\sigma$ meson exchange is less, which is in contrast
with the results of Horowitz et al. [8] who find the contribution
of the $\sigma$ meson exchange (calculated within the Bonn NN potential)
to be larger. On the other hand, the results of
Lee and Riska [7], with meson exchange terms calculated within
the Paris potential, are in agreement with our findings. Of course,
CEOBEM calculations do not involve any explicit dependence on the
NN interaction model. Therefore, they can be used to distinguish
between the predictions of the scalar and vector components of the
heavy meson exchange contributions to the pion production
calculated within the models dependent on NN interaction.

The comparison of our calculations with the data for the total cross sections
for $p(p,p\pi^{0})p$ reaction is shown in Fig. 2. The solid curve
represents the coherent sum of the considered diagramms corresponding
to all the meson exchanges as described above (the relative contributions
of the individual meson exchange graphs remain of the same kind as that
in Fig. 1). The calculations are able to reproduce the absolute magnitude
and beam energy dependence of the measured cross sections very well at the
low as well as higher beam energies in this case too.
Therefore, it is possible to have a consistent
description of both $\pi^{+}$ and $\pi^{0}$ production in $pp$
collisions within COEBEM, which has eluded most of the other theoretical
models so far.

In addition, we show in this figure the
relative contributions of the pion production via excitation
of nucleon (dashed line) and $\Delta$ isobar (dotted line) intermediate
states. Their coherent sum is shown by the solid line.
We notice that non-resonant (nucleon intermediate state)
pion production is predominant at beam energies near the threshold, while
the resonant production becomes important as the beam energy increases.
Thus the near threshold data taken at Bloomington and Uppsala are consistent
with predominantly non-resonant pion production picture.

In Fig. 3, we compare our calculation with the data for the angular
distribution of pions emitted in the reaction $p(p,n\pi^{+})p$ at the
beam energies of 320 MeV, 300 MeV and 294 MeV. The pion angles are in the
$\pi NN$ center-of-mass frame. The solid (dashed) lines show the calculated
cross sections where contributions of both nucleon and delta isobar
(only nucleon) intermediate states are included.
There is a fairly good agreement between the theoretical and
measured angular distributions. At beam energies of 300 MeV and
294 MeV, the contributions of the delta isobar excitation is negligible,
while at 320 MeV delta isobar excitation makes a visible
contribution to the cross section. The near isotropy of the calculated
as well as measured cross sections at beam energies $\leq$ 300 MeV
suggests a $s$-wave pion final state at these energies. However, at 320 MeV
there is some evidence of a non-$s$-wave contribution even in the
non-resonant pion production. This could have consequences on the
polarisation observables.

In summary, the covariant effective one-boson-exchange model with
final state interactions among the outgoing nucleons
taken into account, provides an excellent description of the recently
measured data on $\pi^{+}$ and $\pi^{0}$
production in proton-proton collisions at beam energies near the kinematical
threshold. The $\omega$-meson exchange contributes significantly near
the threshold. In CEOBEM, the heavy meson exchange contributions are
independent of the NN interaction model. They can, thus, serve
to distinguish between various predictions of the scalar and vector
components of the short range axial-charge operator (which has the
same form as that of the non-relativistic $s$-wave pion production) obtained
with different NN interactions. This may have consequences in the
explanation of the large enhancement of the effective axial charge,
found in the analysis of first forbidden $\beta$ transitions in
heavy nuclei [24].  The data on the polarisation observables
will be useful in underlining more details of the nature of the final
state interactions and the neglected pion-rescattering term.

\newpage
\begin{center} {Figure Captions} \end{center}
\begin{itemize}
\item [Fig. 1] The total cross section for the
$p(p,n\pi^{+})p$ reaction as a function of beam energy. The dotted, dashed-
double-dotted, dashed-dotted and dashed curve represent the contributions of
$\rho$, $\sigma$, $\omega$ and $\pi$ meson exchanges respectively. Their
coherent sum is shown by the solid line. The experimental data are from
Ref. [3] and [25].

\item [Fig. 2] The total cross section for the $p(p,p\pi^{0})p$ reaction
as a function of beam energy. The dotted and dashed curves represent the
results of calculations obtained with delta only and nucleon only
intermediate states respectively. Their coherent sum is shown by the
solid line. The experimental data is from Ref. [1] and [25].
The dashed-dotted line shows the results of full calculations obtained
with no final state interaction effects.

\item [Fig. 3] Pion angular distributions in the $\pi NN$ center-of-mass
frame. The dashed lines represent the results of calculations
obtained with nucleon only intermediate states while the solid line is the
total cross sections in which the contributions of nucleon and delta
intermediate states are coherently summed.
\end{itemize}
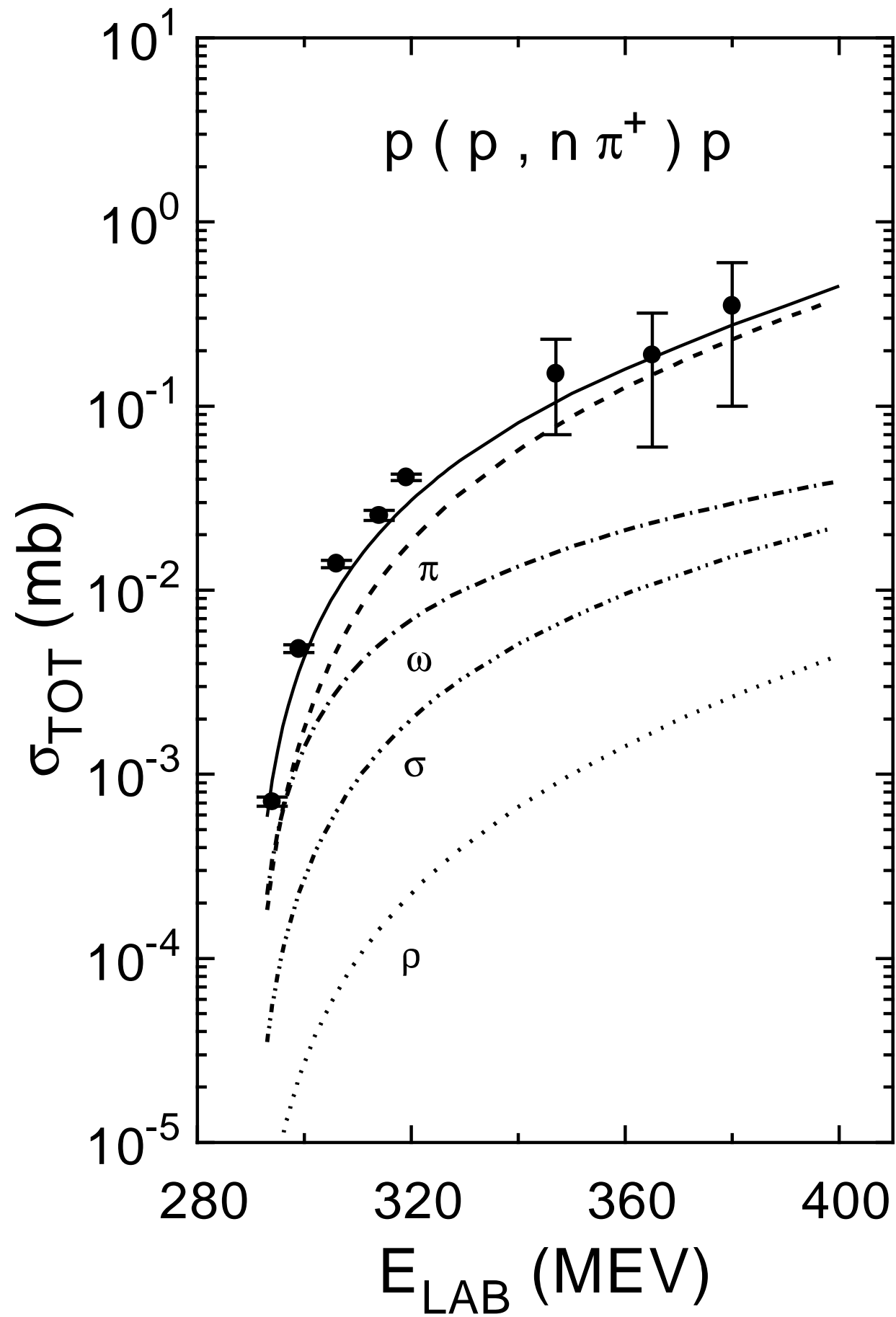
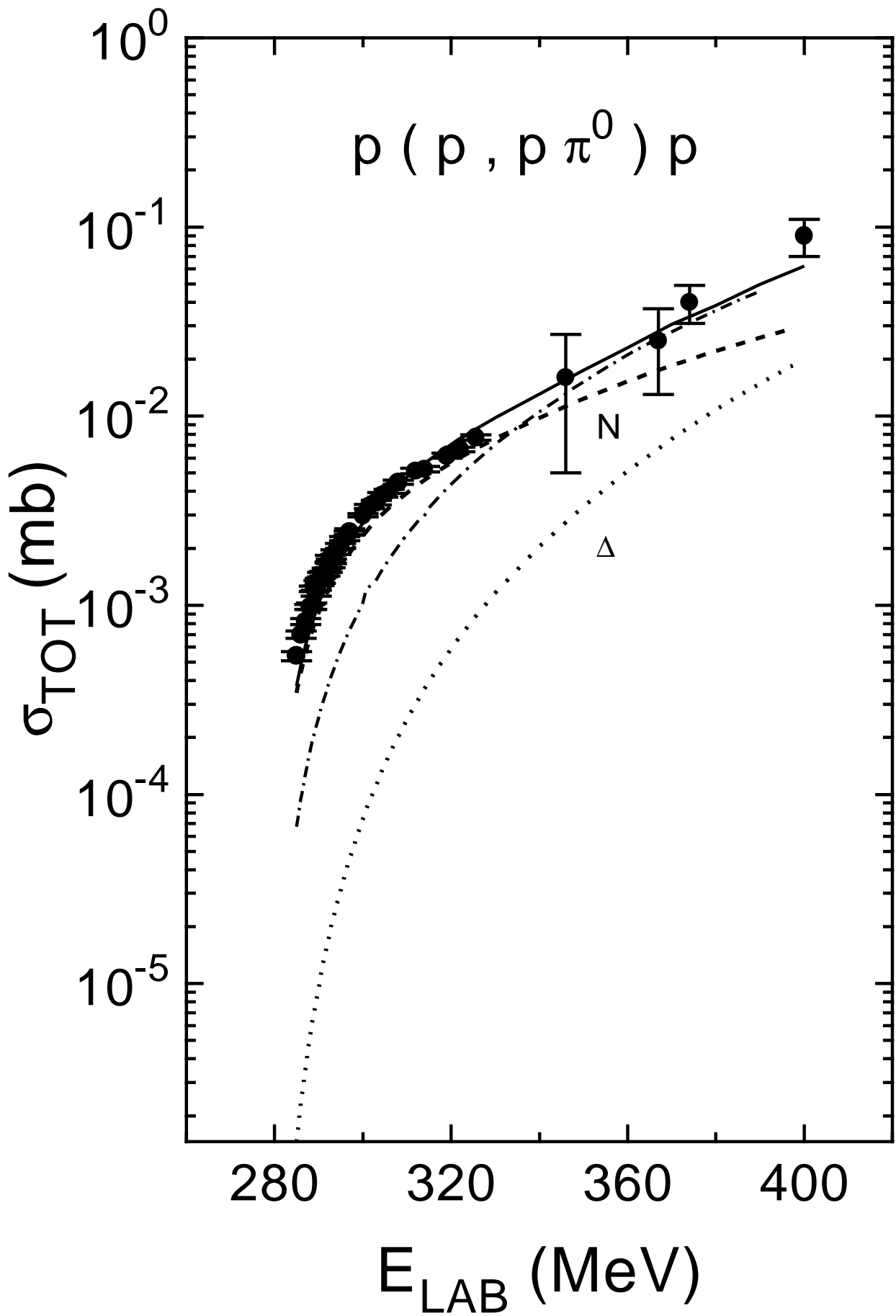
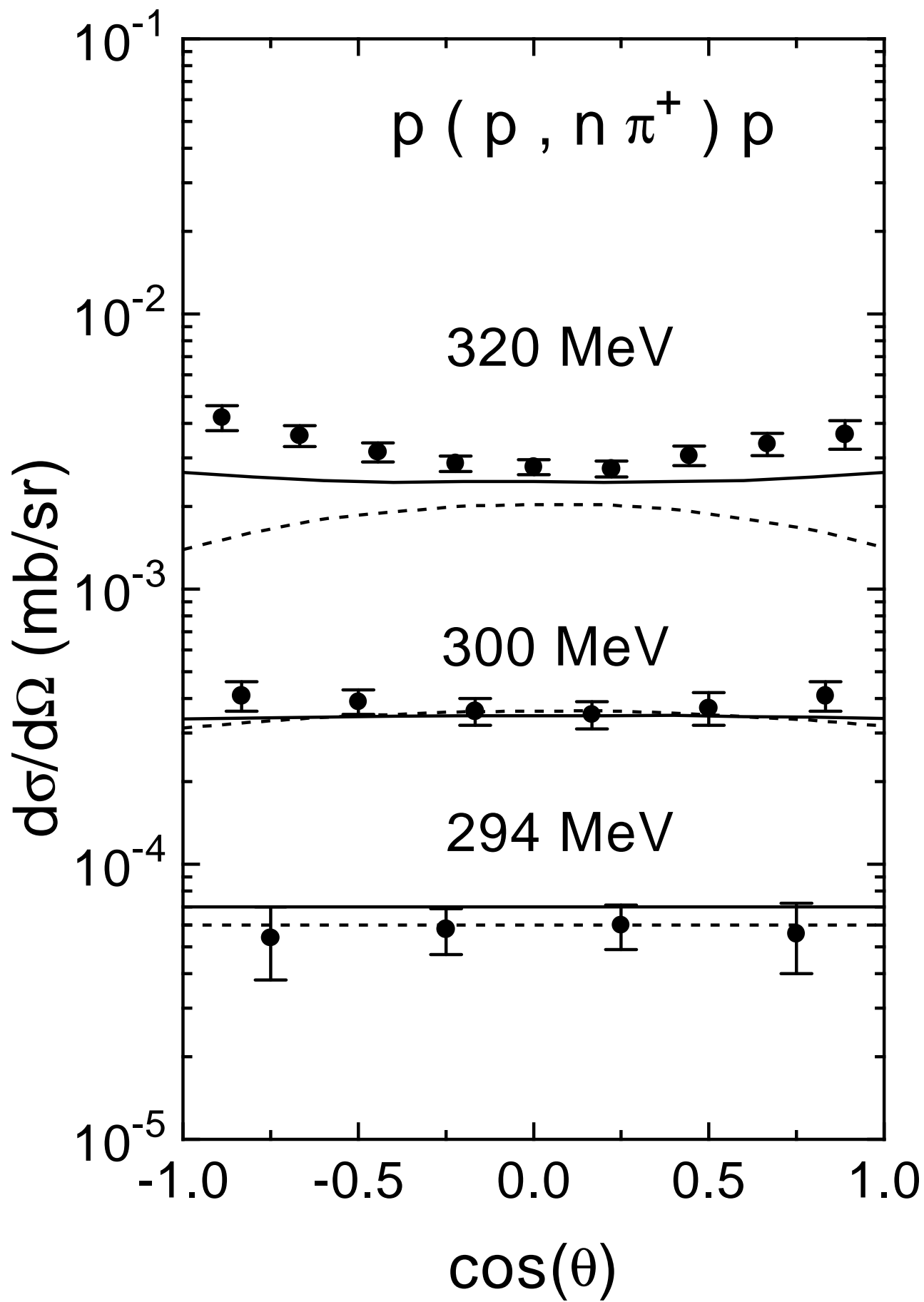
\end{document}